\documentstyle[times,pramana,floats]{ias}
\begin{document}
\ifx\epsfannounce\undefined \def\epsfannounce{\immediate\write16}\fi
 \epsfannounce{This is `epsf.tex' v2.7k <10 July 1997>}%
\newread\epsffilein    
\newif\ifepsfatend     
\newif\ifepsfbbfound   
\newif\ifepsfdraft     
\newif\ifepsffileok    
\newif\ifepsfframe     
\newif\ifepsfshow      
\epsfshowtrue          
\newif\ifepsfshowfilename 
\newif\ifepsfverbose   
\newdimen\epsfframemargin 
\newdimen\epsfframethickness 
\newdimen\epsfrsize    
\newdimen\epsftmp      
\newdimen\epsftsize    
\newdimen\epsfxsize    
\newdimen\epsfysize    
\newdimen\pspoints     
\pspoints = 1bp        
\epsfxsize = 0pt       
\epsfysize = 0pt       
\epsfframemargin = 0pt 
\epsfframethickness = 0.4pt 
\def\epsfbox#1{\global\def\epsfllx{72}\global\def\epsflly{72}%
   \global\def\epsfurx{540}\global\def\epsfury{720}%
   \def\lbracket{[}\def\testit{#1}\ifx\testit\lbracket
   \let\next=\epsfgetlitbb\else\let\next=\epsfnormal\fi\next{#1}}%
%
%
\def\epsfgetlitbb#1#2 #3 #4 #5]#6{%
   \epsfgrab #2 #3 #4 #5 .\\%
   \epsfsetsize
   \epsfstatus{#6}%
   \epsfsetgraph{#6}%
}%
\def\epsfnormal#1{%
    \epsfgetbb{#1}%
    \epsfsetgraph{#1}%
}%
\newhelp\epsfnoopenhelp{The PostScript image file must be findable by
TeX, i.e., somewhere in the TEXINPUTS (or equivalent) path.}%
\def\epsfgetbb#1{%
%
%
    \openin\epsffilein=#1
    \ifeof\epsffilein
        \errhelp = \epsfnoopenhelp
        \errmessage{Could not open file #1, ignoring it}%
    \else                       
        {
            \chardef\other=12
            \def\do##1{\catcode`##1=\other}%
            \dospecials
            \catcode`\ =10
            \epsffileoktrue         
            \epsfatendfalse     
            \loop               
                \read\epsffilein to \epsffileline
                \ifeof\epsffilein 
                \epsffileokfalse 
            \else                
                \expandafter\epsfaux\epsffileline:. \\%
            \fi
            \ifepsffileok
            \repeat
            \ifepsfbbfound
            \else
                \ifepsfverbose
                    \immediate\write16{No BoundingBox comment found in %
                                    file #1; using defaults}%
                \fi
            \fi
        }
        \closein\epsffilein
    \fi                         
    \epsfsetsize                
    \epsfstatus{#1}%
}%
%
\def\epsfclipon{\def\epsfclipstring{ clip}}%
\def\epsfclipoff{\def\epsfclipstring{\ifepsfdraft\space clip\fi}}%
\epsfclipoff 
%
%
\def\epsfspecial#1{%
     \epsftmp=10\epsfxsize
     \divide\epsftmp\pspoints
     \ifnum\epsfrsize=0\relax
       \includegraphics{\ifepsfdraft}%
     \else
       \epsfrsize=10\epsfysize
       \divide\epsfrsize\pspoints
       \includegraphics{\ifepsfdraft}%
     \fi
}%
%
\def\epsfframe#1%
{%
  \leavevmode                   
  \setbox0 = \hbox{#1}%
  \dimen0 = \wd0                                
  \advance \dimen0 by 2\epsfframemargin         
  \advance \dimen0 by 2\epsfframethickness      
  \vbox
  {%
    \hrule height \epsfframethickness depth 0pt
    \hbox to \dimen0
    {%
      \hss
      \vrule width \epsfframethickness
      \kern \epsfframemargin
      \vbox {\kern \epsfframemargin \box0 \kern \epsfframemargin }%
      \kern \epsfframemargin
      \vrule width \epsfframethickness
      \hss
    }
    \hrule height 0pt depth \epsfframethickness
  }
}%
\def\epsfsetgraph#1%
{%
   %
   %
   \leavevmode
   \hbox{
     \ifepsfframe\expandafter\epsfframe\fi
     {\vbox to\epsfysize
     {%
        \ifepsfshow
            \vfil
            \hbox to \epsfxsize{\epsfspecial{#1}\hfil}%
        \else
            \vfil
            \hbox to\epsfxsize{%
               \hss
               \ifepsfshowfilename
               {%
                  \epsfframemargin=3pt 
                  \epsfframe{{\tt #1}}%
               }%
               \fi
               \hss
            }%
            \vfil
        \fi
     }%
   }}%
   %
   %
   \global\epsfxsize=0pt
   \global\epsfysize=0pt
}%
%
%
\def\epsfsetsize
{%
   \epsfrsize=\epsfury\pspoints
   \advance\epsfrsize by-\epsflly\pspoints
   \epsftsize=\epsfurx\pspoints
   \advance\epsftsize by-\epsfllx\pspoints
%
%
   \epsfxsize=\epsfsize{\epsftsize}{\epsfrsize}%
   \ifnum \epsfxsize=0
      \ifnum \epsfysize=0
        \epsfxsize=\epsftsize
        \epsfysize=\epsfrsize
        \epsfrsize=0pt
%
%
      \else
        \epsftmp=\epsftsize \divide\epsftmp\epsfrsize
        \epsfxsize=\epsfysize \multiply\epsfxsize\epsftmp
        \multiply\epsftmp\epsfrsize \advance\epsftsize-\epsftmp
        \epsftmp=\epsfysize
        \loop \advance\epsftsize\epsftsize \divide\epsftmp 2
        \ifnum \epsftmp>0
           \ifnum \epsftsize<\epsfrsize
           \else
              \advance\epsftsize-\epsfrsize \advance\epsfxsize\epsftmp
           \fi
        \repeat
        \epsfrsize=0pt
      \fi
   \else
     \ifnum \epsfysize=0
       \epsftmp=\epsfrsize \divide\epsftmp\epsftsize
       \epsfysize=\epsfxsize \multiply\epsfysize\epsftmp
       \multiply\epsftmp\epsftsize \advance\epsfrsize-\epsftmp
       \epsftmp=\epsfxsize
       \loop \advance\epsfrsize\epsfrsize \divide\epsftmp 2
       \ifnum \epsftmp>0
          \ifnum \epsfrsize<\epsftsize
          \else
             \advance\epsfrsize-\epsftsize \advance\epsfysize\epsftmp
          \fi
       \repeat
       \epsfrsize=0pt
     \else
       \epsfrsize=\epsfysize
     \fi
   \fi
}%
%
%
\def\epsfstatus#1{
   \ifepsfverbose
     \immediate\write16{#1: BoundingBox:
                  llx = \epsfllx\space lly = \epsflly\space
                  urx = \epsfurx\space ury = \epsfury\space}%
     \immediate\write16{#1: scaled width = \the\epsfxsize\space
                  scaled height = \the\epsfysize}%
   \fi
}%
%
%
{\catcode`\%=12 \global\let\epsfpercent=
\global\def\epsfatend{(atend)}%
%
%
%
%
%
%
%
\long\def\epsfaux#1#2:#3\\%
{%
   \def\testit{#2}
   \ifx#1\epsfpercent           
       \ifx\testit\epsfbblit    
            \epsfgrab #3 . . . \\%
            \ifx\epsfllx\epsfatend 
                \global\epsfatendtrue
            \else               
                \ifepsfatend    
                \else           
                    \epsffileokfalse
                \fi
                \global\epsfbbfoundtrue
            \fi
       \fi
   \fi
}%
%
%
\def\epsfempty{}%
\def\epsfgrab #1 #2 #3 #4 #5\\{%
   \global\def\epsfllx{#1}\ifx\epsfllx\epsfempty
      \epsfgrab #2 #3 #4 #5 .\\\else
   \global\def\epsflly{#2}%
   \global\def\epsfurx{#3}\global\def\epsfury{#4}\fi
}%
%
%
\def\epsfsize#1#2{\epsfxsize}%
%
%
\let\epsffile=\epsfbox

\epsfclipon
\def\sfrac#1#2{{\textstyle\frac{#1}{#2}}}
\newcommand{\beq}{\begin{equation}}
\newcommand{\eeq}{\end{equation}}
\newcommand{\bea}{\begin{eqnarray}}
\newcommand{\eea}{\end{eqnarray}}

\newcommand{\lsim}{\mbox{\raisebox{-.6ex}{~$\stackrel{<}{\sim}$~}}}
\newcommand{\gsim}{\mbox{\raisebox{-.6ex}{~$\stackrel{>}{\sim}$~}}}
\title{Inflation, Large Scale Structure and Particle Physics}

\author{S.F.King}
\address{Department of Physics and Astronomy, University of
Southampton, Southampton SO17 1BJ, U.K.} \keywords{inflation}
\abstract{We review experimental and theoretical developments in
inflation and its application to structure formation, including the curvaton idea.
We then discuss a particle physics model of supersymmetric hybrid inflation
at the intermediate scale in which the Higgs scalar field is responsible
for large scale structure, show how such a theory is completely natural
in the framework extra dimensions with an intermediate string scale.}

\maketitle
\section{Introduction}
The Standard Hot Big Bang (SHBB) model of cosmology provides a
convincing description of the early universe, from the time of
nucleosynthesis, when the universe was a few seconds old, to the
decoupling time 380,000 years later when the electrons were bound
into atoms and the universe became transparent. The radiation that
was emitted from the last scattering surface at this time is
observed today 13.7 billion years later, redshifted by a factor of
$z\approx 1090$ as the cosmic microwave background (CMB). The
homogeneity and isotropy of the CMB implies very simple initial
conditions for the early universe. The temperature fluctuations in
the CMB, observed initially by COBE, have been measured most
recently to high precision by WMAP as shown in Figure \ref{map}
\cite{WMAP}. A compilation of data on the angular power spectrum
is shown in the left panel of Figure 2, and a best fit
curve to these data is shown also shown in the right panel of this
Figure.

\begin{figure}[b]
\centerline{\epsfxsize=0.7\textwidth\epsfbox{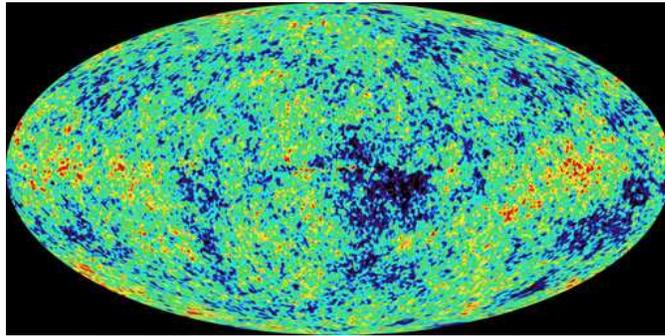}} \caption{The
CMB sky as seen by WMAP, showing the anisotropies at a few parts
in $10^5$.} \label{map}
\end{figure}

\begin{figure}
\centerline{\epsfxsize=0.5\textwidth\epsfbox{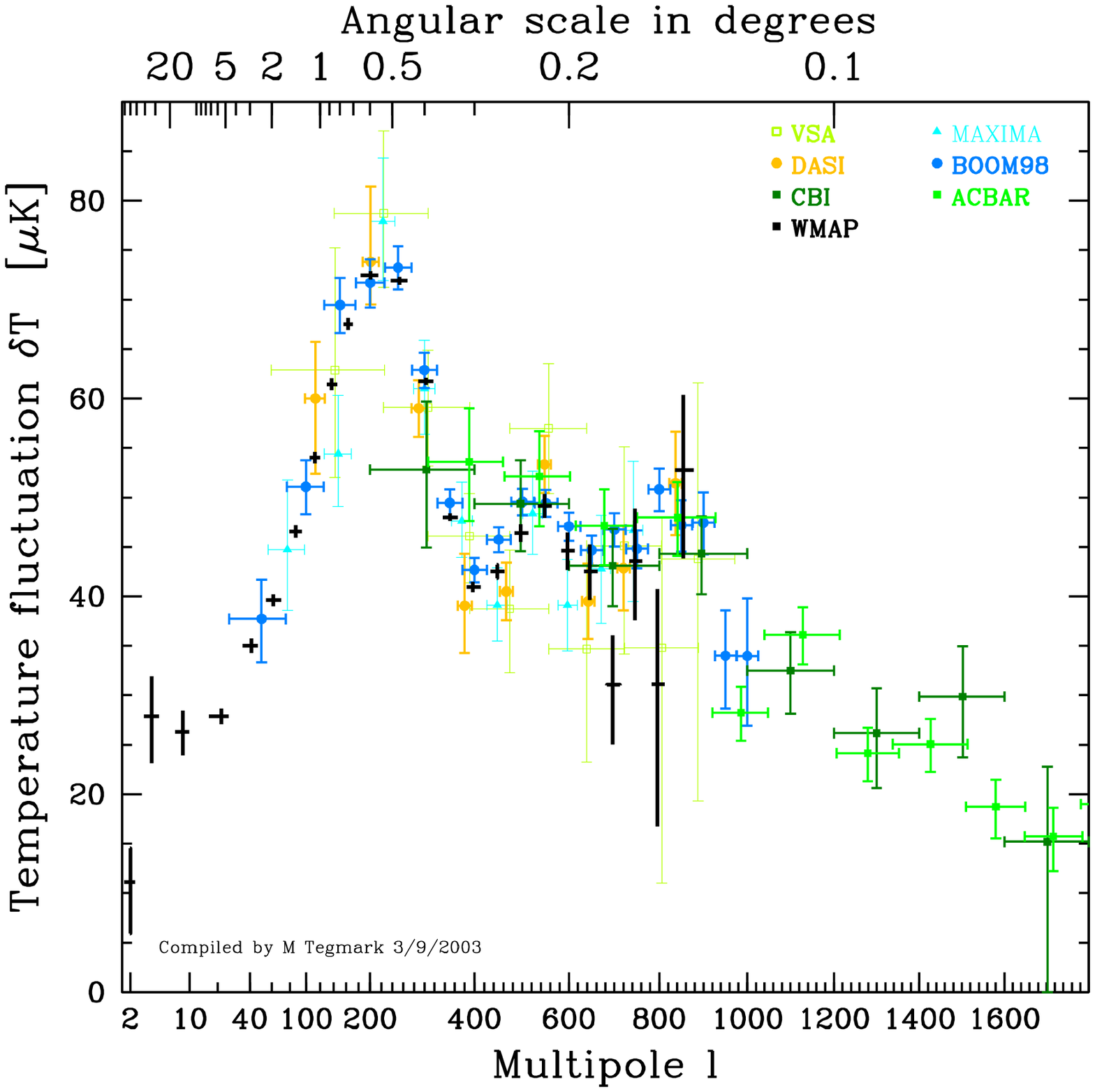}\epsfxsize=0.5\textwidth\epsfbox{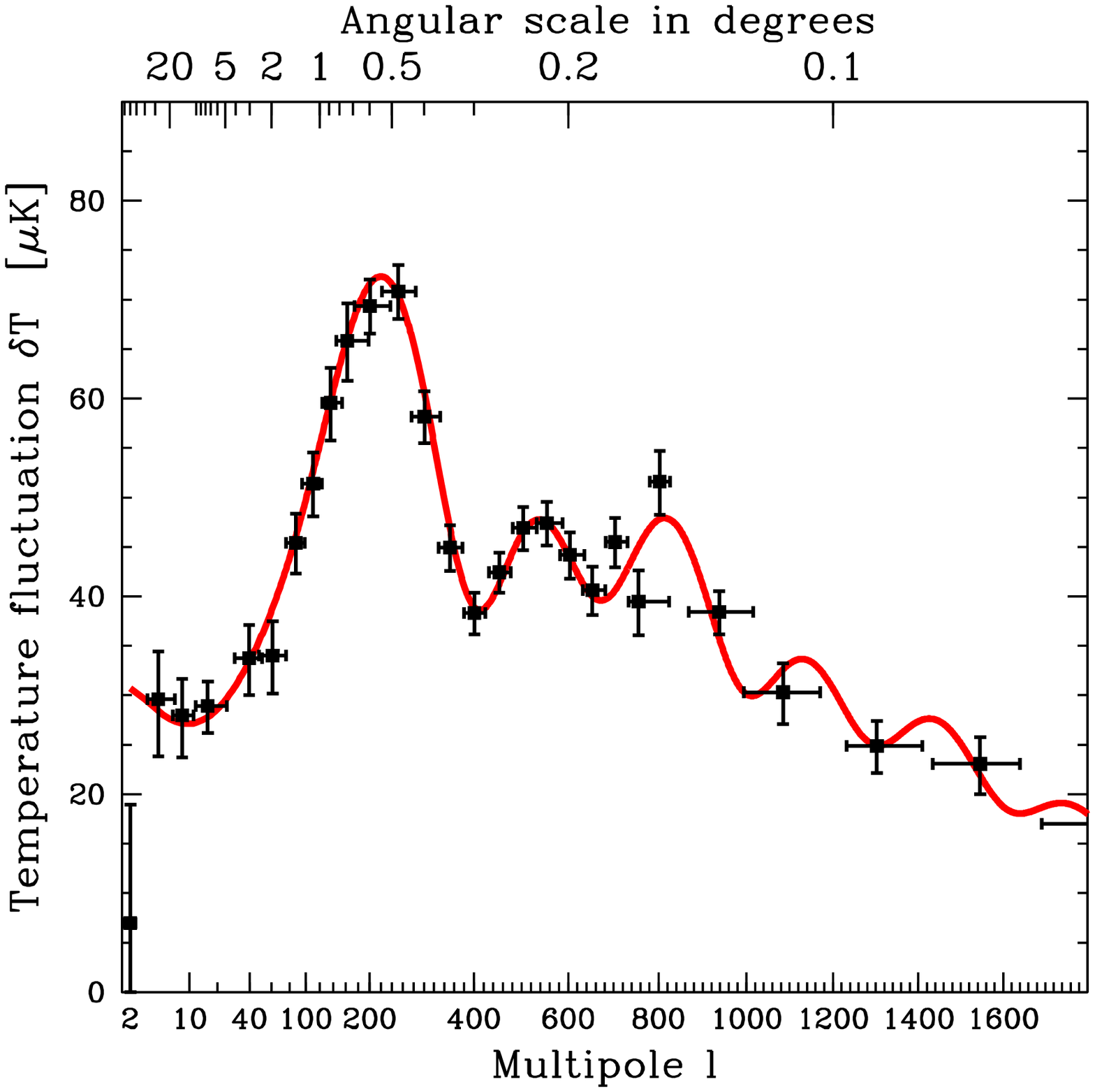}}
\caption{The angular power spectrum compiled from the most recent
experiments (left), and the best fit theoretical curve (right)
corresponding to a $\Lambda CDM$ universe as discussed in the
text.} \label{exp}
\end{figure}

The position of the first peak of the angular power spectrum shows
that the universe is flat to within experimental error, and the
detailed fit to these data show that the necessary ingredients in
the universe, must include about 23\% cold dark matter (CDM) and
73\% dark energy (DE), with baryons only providing 4\% of the
matter in the universe, of which only a tenth form luminous stars.
The observed temperature fluctuations, together with DM and DE,
seems sufficient to give the observed clumping of stars into
galaxies, galaxies into clusters and clusters into superclusters,
reproducing the large scale structure in the universe.
\footnote{Incidentally, hot dark matter (HDM) in the form of
neutrinos is disfavoured, since it would tend to wash out the
observed galaxy structures on small scales. This leads to an upper
bound on the sum of masses of the three neutrinos, which when
combined with the neutrino oscillation data implies an upper bound
on each neutrino flavour of about 0.23 eV \cite{Pierce:2003uh}.}

\begin{figure}
\centerline{\epsfxsize=0.6\textwidth\epsfbox{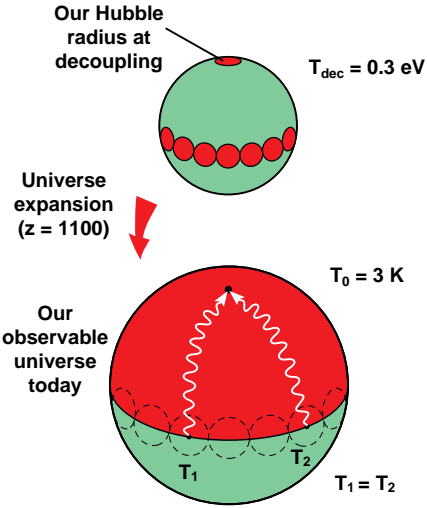} }
\caption{A depiction of the horizon problem (from [3]).}
\label{hom}
\end{figure}

From the point of view of the SHBB the current paradigm of our
universe as described above presents several puzzles. The
homogeneity and isotropy of the CMB is a major puzzle, since,
according to the SHBB, the radiation from points of the surface of
last photon scattering which are more than 1 degree apart could
never have been in causal contact, as depicted in Figure \ref{hom}
(taken from \cite{Garcia-Bellido:2003ih}). This well known puzzle
is called the horizon problem. Another puzzle from the point of
view of the SHBB is the fact that the universe is approximately
flat, since devations in the early universe from perfect flatness
would have by now become magnified. The fact that the universe
today is accurately flat only makes this so called flatness
problem even more puzzling. The temperature fluctuations of the
CMB, although required as seeds of structure formation, also admit
no explanation in the SHBB framework. There are additional puzzles
associated with the matter and energy content of the universe,
namely why is the baryon to photon ratio is a number of order
$10^{-10}$ and not zero? What is the origin of CDM?  What is the
source of DE and why should its density be of the same order as
that of CDM at the present epoch?

Inflation is an attempt to solve the horizon and flatness
problems, which also incorporates a quantum origin of temperature
fluctuations. Moreover the basic physics of inflation is not too
dissimilar from the behaviour of the present day universe, namely
accelerated expansion due to vacuum energy. The original proposal
of Guth in 1981 \cite{Guth:1980zm}, posited that the energy
density of the early universe was dominated by some scalar field
which became hung up in a false vacuum for some time during which
the energy density of the universe was approximately constant
leading to approximately exponential expansion of space, thereby
solving the horizon and flatness problems. Such inflation also
effectively dilutes any unwanted cosmological relics produced at
an earlier epoch such as magnetic monopoles predicted by certain
gauge unified theories. The original version of inflation
described above does not work since the false-vacuum tunnels to
true vacuum due to a first order phase transition, and bubbles of
true vacuum grow due to negative pressure, leading to a never
ending sequence of inflating universes. However soon after this
proposal versions of inflation were proposed
\cite{Linde:1981mu,Albrecht:1982wi} based on an inflaton field
with a very flat potential higher than its minimum, along which
the field slowly rolls during which the universe inflates, before
reaching the true ground state about which the inflaton field
oscillates, ending inflation and reheating the universe as it
decays to the minimum. This scenario called slow roll inflation is
depicted in Figure \ref{slow} (taken from
\cite{Garcia-Bellido:2003ih}).

\begin{figure}
\centerline{\epsfxsize=0.7\textwidth\epsfbox{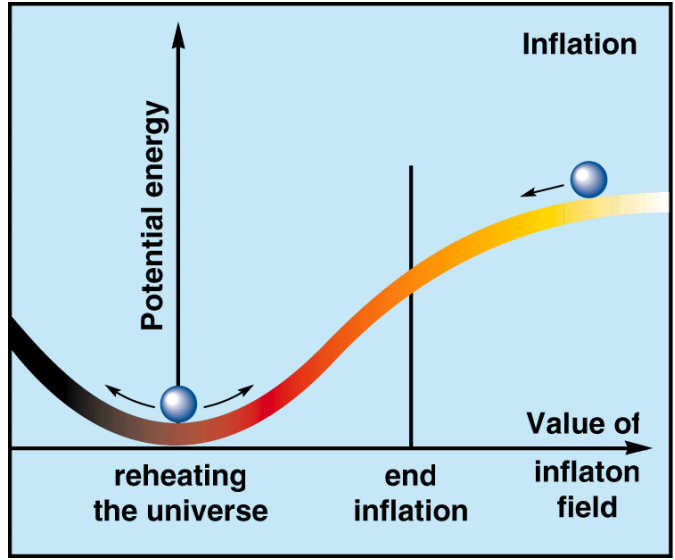} }
\caption{A depiction of slow roll inflation (from [3]).}
\label{slow}
\end{figure}

In such an inflationary approach, the very largest scales, which
are now entering the horizon, would have been in causal contact at
very early times, thereby solving the horizon problem, as shown in
Figure \ref{inf4}. It also accounts for the observed flatness of
the universe $\Omega =1$ (the flatness problem), consistent with
the CMB data. \footnote{Alternative versions of inflation with
$\Omega>1$ are fine-tuned \cite{Linde:2003hc}.}

\begin{figure}[h]
\centerline{\epsfxsize=0.7\textwidth\epsfbox{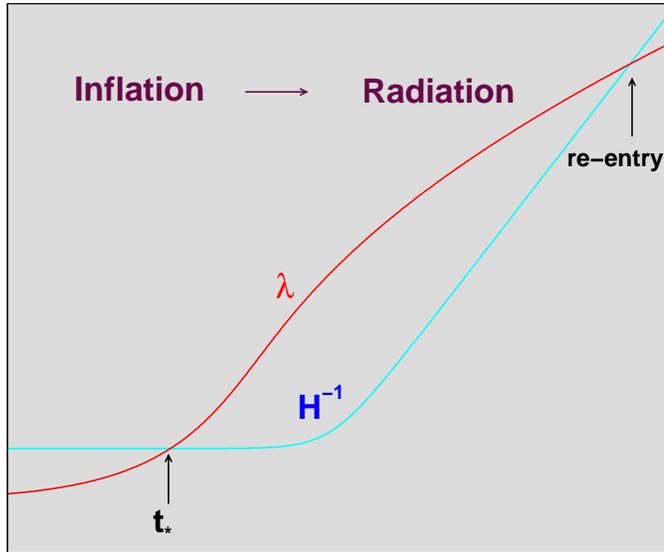} }
\caption{How inflation solves the horizon problem. Scales
$\lambda$ go out of causal contact at the time of horizon exit
$t_*$, and re-enter the horizon much later.} \label{inf4}
\end{figure}

During the epoch of slow roll inflation in Figure \ref{slow},
the equation of motion for the field during the inflationary phase
is given by
\beq
\ddot{\phi}+3H\dot{\phi}=-V'
\eeq
where the Hubble constant in this epoch is given by
\beq
H^2\approx \frac{V}{2m_P^2}
\eeq
where $m_P$ is the reduced Planck mass.
The conventionally defined slow roll parameters are
\begin{eqnarray}
 \epsilon &\equiv& \frac{1}{2} m_P^2
  \left( \frac{V'}{V} \right)^{2} \ll 1  \label{eq:slowepsilon} \\
 | \eta | &\equiv& \left| \frac{ m_P^2 V''}{V} \right| \ll 1
  \label{eq:sloweta}
\end{eqnarray}
where $V' (V'')$ are the first (second) derivatives of the potential.
Assuming the slow roll conditions are satisfied, the field equation
approximates to
\beq
3H\dot{\phi}=-V'.
\label{field}
\eeq

The current status of inflation is summarized in Table 1.
\begin{table}[h]
\caption{Table 1. Current Status of Inflation.}
\hskip4pc\vbox{\columnwidth=26pc
\begin{tabular}{ll}
Prediction & Observation  \\ \hline
$\Omega_0=1$  & $\Omega_0=1.0\pm 0.02$ \\
Density perturbations  & Observed perturbations\\
1.Acoustic peaks expected & 1.Three acoustic peaks observed\\
2.Gaussian spectrum & 2.No evidence for non-Gaussianity\\
3.Spectral index $n_s\approx 1$ & 3. One sigma range $n_s=0.94-1.02$\\
4.Gravity waves $\epsilon \ll 1$ & 4.Gravity waves not observed
$\epsilon < 0.022$
\end{tabular}
}
\end{table}

\section{Hybrid Inflation}

There are essentially three types of slow roll inflation which can
be categorised as large field (chaotic) inflation, small field
(new) inflation, and hybrid inflation. From the point of view of
particle physics models perhaps the most interesting of these is
hybrid inflation \cite{Bento:1991ax,Copeland:1994vg} which involves two (or
more) scalar fields: a slowly rolling inflaton field $\phi$ plus
other fields $N,\ldots $ which are held at zero (or small) field
values during inflation, but which are destabilised when $\phi$
reaches a critical value called $\phi_c$. In hybrid inflation,
depicted in Figure \ref{infold}, during inflation only the
inflaton field $\phi$ non-zero, and the potential takes an
extremely simple form 
\beq V=V_0+\frac{1}{2}m^2\phi^2 \eeq 
where
the idea is that the mass term $\frac{1}{2}m^2\phi^2$ is much
smaller than the constant energy density term $V_0$. The Hubble
constant is given by $H\approx \frac{V_0^{1/2}}{\sqrt{3}m_P}$. 

\begin{figure}
\centerline{\epsfxsize=0.7\textwidth\epsfbox{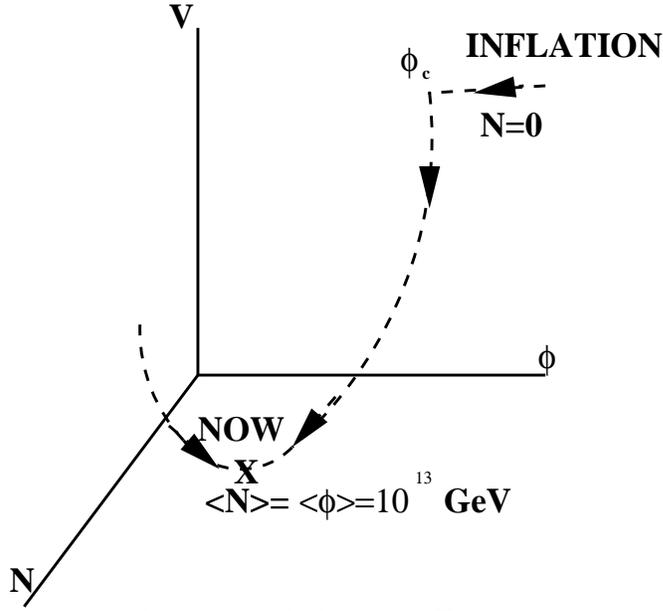} }
\caption{A depiction of hybrid inflation. The parameters at the
global minimum correspond to the specific model discussed in the
text.} \label{infold}
\end{figure}

The slow roll parameters during inflation are given by
\beq
\epsilon \approx \frac{m_P^2}{2}\left(\frac{m^2\phi^2}{V_0}\right)^2, \ \ \ \
\eta  \approx  \frac{m_P^2m^2}{V_0}
\label{hybridslow}
\eeq
Assuming the slow roll conditions are satisfied $\epsilon,\eta \ll 1$
the field equation gives $\dot{\phi}\approx -\frac{m^2\phi}{3H}$.

To understand the origin of the critical value of $\phi$ which ends
inflation, one must write down the full potential of the theory.
This will be model dependent but will include a term like
$(\phi^2-\phi_c^2)N^2$ such that when $\phi >\phi_c$ the term is
positive, corresponding to a positive mass squared for the $N$ field,
which holds this field st $N=0$, but when
$\phi <\phi_c$ the term becomes
negative, corresponding to a negative (tachyonic)
mass squared for the $N$ field, which causes it to become non-zero,
effectively ending inflation, as shown in Figure \ref{infold}.

An interesting feature of hybrid inflation is that, since
$\phi <m_P$, we have $\epsilon \ll \eta $ as is clear from
Eq.\ref{hybridslow}. Since the parameter $\epsilon$ is
responsible for the tensor modes associated with cosmological
gravitational waves, in hybrid inflation gravitational waves
are predicted to be a negligible effect.

The ratio of tensor to scalar amplitudes is conventionally defined
to be $R$, and the scalar and tensor spectral indices
$n_s,n_t$ are then given in terms of the slow roll parameters as
\beq
R \approx   -8n_t\approx 16 \epsilon, \ \ \ \
n_s \approx 1+2\eta - 6\epsilon 
\eeq

The current experimental
one and three $\sigma$ limits on $R,n_s$ are shown in Figure
\ref{barger} (taken from \cite{Barger:2003ym}), together with the
theoretical expectations for different types of inflation. Since
the point $R=0, n_s=1$ is permitted, it is clear that the present
data does not discriminate between the different slow roll
inflationary models.

\begin{figure}[h]
\centerline{\epsfxsize=0.7\textwidth\epsfbox{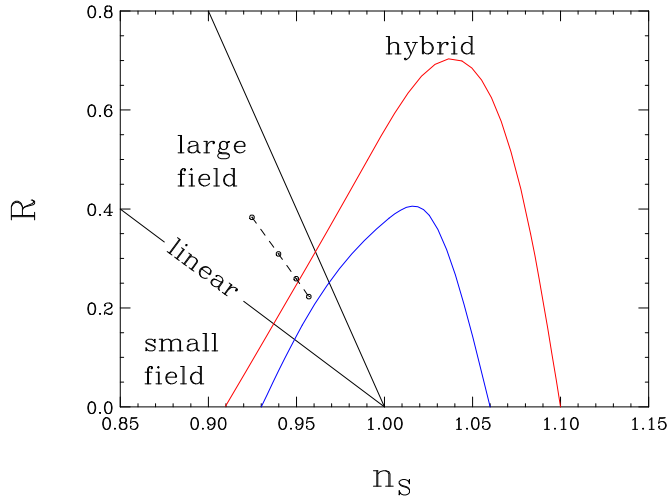}}
\caption{Experimental limits at one and three sigma in the (R,$n_s$)
plane, as compared to the expectations from different types of
inflation model (from [9]).} \label{barger}
\end{figure}

\section{Structure Formation}

Another commonly stated success of inflation is the fact that the
observed primordial density perturbations, which were first
observed by COBE on cosmological scales just entering the horizon,
and which are supposed to be the seeds of large scale structure,
could have originated from the quantum fluctuations of the
inflaton field, the scalar field which is supposed to be
responsible for driving inflation. In this scenario the quantum
fluctuations of the inflaton field during the period of inflation
become classical perturbations at horizon exit, giving a
primordial curvature perturbation which remains constant until the
approach of horizon entry.

\begin{figure}[h]
\centerline{\epsfxsize=0.5\textwidth\epsfbox{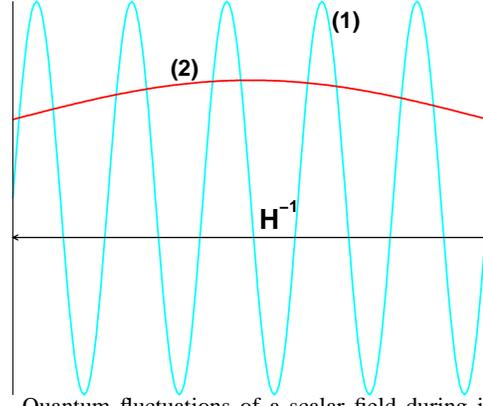} }
\caption{Quantum fluctuations of a scalar field during inflation
different may be Fourier decomposed and behave differently
according to their wavelength. The modes (1) shorter than the
horizon size become redshifted, while the modes (2) longer than
the horizon size get frozen in (figure from Mar Bastero-Gil).} \label{inf3}
\end{figure}

The equation for the fluctuations of the inflaton field is \beq
\delta \ddot{\phi} +3H\delta \dot{\phi} - \nabla^2\delta\phi =0
\eeq The field fluctuations $\delta \phi$ may then be Fourier
decomposed so that $\delta \phi_k$ refers to the fluctuations of
the field associated with a particular mode. The spectrum of the
fluctuations are then defined as \beq P_\phi \equiv
\frac{k^3}{2\pi^2}<|\delta \phi_k|^2>=
\left(\frac{H_*}{2\pi}\right)^2 \label{fluc} \eeq where
$k=2\pi/\lambda$ refers to the wavenumber of the particular mode
of wavelength $\lambda$, and $*$ denotes quantities evaluated at
the time of horizon exit. For wavelengths within the Hubble
horizon distance $\lambda <H^{-1}$ the field fluctuations become
redshifted as $\delta \phi_k\sim a^{-1}$, as depicted in Figure \ref{inf3}.
While for wavelengths
greater than the Hubble horizon distance $\lambda >H^{-1}$ the
fluctuation become frozen in and are given from Eq.\ref{fluc} as
$\delta \phi_k(t_*)\sim H_*/\sqrt{2k^3}$.

The fluctuations
$\delta \phi_k(t_*)$ are then frozen in as a classical
perturbation on the density and hence the curvature perturbation
$\cal{R}$, given by
\beq
{\cal R}=-\frac{H_*}{\dot{\phi}_*}\delta
\phi_k= -\frac{H_*}{\dot{\phi}_*}\frac{H_*}{\sqrt{2k^3}} \label{R0}
\eeq
The primordial power spectrum, the seed of the observed
linear power spectrum $P(k)$ is then given by $P_{{\cal R}}
\equiv \frac{k^3}{2\pi^2}<|{\cal R}|^2>$.
Experimentally the primordial power spectrum is observed to be
approximately scale invariant
\beq
P_{{\cal R}}(k) =
A_s\left(\frac{k}{k_*}\right)^{n_s-1}\approx A_s \label{inv}
\eeq
where $A_s$ is the scalar amplitude, and $n_s$ is the spectral
index which is observed to be approximately scale invariant and
approximately equal to unity. From Eq.\ref{R} we find
\beq
P_{{\cal R}}^{1/2} \approx \frac{H_*^2}{2\pi\dot{\phi}_*}
\approx \frac{1}{2\pi\sqrt{3}}\frac{V^{3/2}}{m_P^3V'}
\label{theory}
\eeq
The theoretical normalisation in
Eq.\ref{theory}, evaluated from the inflaton potential, must then
be compared to the experimentally measured normalisation from WMAP
\beq
P_{{\cal R}}^{1/2} \approx (4-5)\times 10^{-5} \label{exp1}
\eeq
The comparison of Eqs.\ref{theory} to \ref{exp1} provides a
very strong restriction on the inflaton potential, much stronger
than the requirements of slow roll.

\section{The Curvaton}

The advantage of the scenario in which the inflaton fluctuations
are responsible for large scale structure
is that the prediction for the
nearly scale-invariant spectrum depends only on the form of the
inflaton potential, and is independent of what goes on between the
end of inflation and horizon entry. The disadvantage is that it
provides a strong restriction on models of inflation. The price of
such simplicity, with one field being responsible for both
inflation and the primordial curvature perturbation often
translates into a severe restriction on the parameters of the
inflaton potential. This often requires very small values for the
couplings and/or the masses which apparently renders many such
theories unnatural.

Recently it has been pointed out that in general it is unnecessary
for the inflaton field to be responsible for generating the
curvature perturbation
\cite{Lyth:2001nq,Lyth:2002my,Moroi:2001ct,Moroi:2002rd,Dimopoulos:2002kt,Dimopoulos:2003ii,Dimopoulos:2003az}.
It is possible that the inflaton only generates a very small
curvature perturbation during the period of inflation,
which instead may result from the
isocurvature perturbations of a curvaton field
which subsequently become converted into
curvature perturbations in the period after inflation, but before
horizon entry.
Isocurvature perturbations simply mean perturbations which
do not perturb the total curvature, usually because the curvaton
field contributes a very small energy density $\rho_{\sigma}$
during inflation. In the
scenarios presented so far
\cite{Lyth:2001nq,Lyth:2002my,Moroi:2001ct,Moroi:2002rd,Dimopoulos:2002kt,Dimopoulos:2003ii,Dimopoulos:2003az},
the curvaton is assumed to be
completely decoupled from inflationary dynamics, and
is assumed to be some late-decaying scalar which decays before the
time that neutrinos become decoupled.

The reason why the curvaton is assumed to be late-decaying can be
understood from the following argument.
After reheating
the total curvature perturbation can be written as,
\beq
{\cal R} =  (1-f){\cal R}_{\rm r}+f{\cal R}_{\sigma},
\ \ \ \ f=\frac{3\rho_{\sigma}}{4\rho_{\rm r}+3\rho_{\sigma}}
\label{R}
\eeq
and (on unperturbed hypersurfaces on super-horizon sized scales)
\beq
{\cal R}_i\approx -H\left(\frac{\delta \rho_i}{\dot{\rho}_i}\right)
\sim \left(\frac{\delta \rho_i}{{\rho}_i}\right)
\label{Ri}
\eeq
where the curvaton density $\rho_{\sigma}$
and radiation density $\rho_{r}$, arising from the decay of the inflaton,
each satisfy their own
energy conservation equations and each ${\cal R}_i$ remains constant
on super-horizon scales. The time evolution of ${\cal R}$
on these scales is then given by its time derivative,
\beq
\dot{\cal R}\approx -H f(1-f)\frac{S_{\sigma {\rm r}}}{3}, \ \ \ \
S_{\sigma {\rm r}} \simeq -3 ({\cal R}_{\sigma}-{\cal R}_{\rm r})
\label{Rdot}
\eeq
where $S_{\sigma {\rm r}}$ is the entropy perturbation.
The curvaton generates an isocurvature perturbation because
initially $\rho_{r}\gg \rho_{\sigma}$, and hence $f\ll 1$,
so that from Eq. (\ref{R}) the curvature perturbation
is dominated by ${\cal R}_{\rm r}$.
However as the universe expands and the scale factor $a$ increases
while the Hubble constant $H$ decreases, the curvaton with mass $m$,
whose oscillations have effectively been frozen in by the large Hubble
constant, begins to oscillate and act as matter.
After this happens $\rho_{r}$ decreases as $a^{-4}$ while the
energy density in the curvaton field $\rho_{\sigma}$ has a slower
fall-off as $a^{-3}$. Eventually the curvaton energy density $\rho_{\sigma}$
becomes comparable to the radiation density from the inflaton decay
$\rho_{r}$, and when this happens we see that $f\sim 1$
and from Eq. (\ref{Rdot}) this
leads to the growth of the total curvature perturbation
${\cal R}$ from the isocurvature perturbation
${\cal R}_{\sigma}>{\cal R}_{\rm r}$.
This mechanism, which allows the curvaton isocurvature perturbations to become
converted into the total curvature perturbation, requires
the curvaton scalar to be late-decaying.

\section{A Particle Physics Model of Inflation and Structure}

We now turn to the problem of constructing a particle physics model of hybrid inflation.
Such a model should be supersymmetric, because supersymmetry has flat directions which make
good candidates for slow roll inflation. On the other hand supersymmetry is necessarily broken in the
inflationary epoch
by the F or D-term vevs which is necessary to generate the vacuum energy $V(0)$ and drive inflation.
Nevertheless supersymmetry remains the best candidate for maintaining a flat potential
and for safeguarding its flatness from radiative corrections in a controllable way.
Assuming supersymmetry, the model must then provide an origin of the supersymmetric
Higgs mass $\mu$, and ideally should also provide a solution to the strong CP problem
via the Peccei-Quinn mechanism. Hybrid inflation models which also include grand unification
typically face problems with magnetic monopoles \cite{King:1997ia}. This is ironic since
the avoidance of monopoles was one of Guth's main motivations for introducing inflation.
Following Guth's philosophy inflation can provide the solution to this problem providing
inflation takes place at a scale below the GUT scale, when inflation will dilute the
monopole abundance.
To avoid the monopole problem, and shed light on the
strong CP problem, and $\mu$ problem, we shall therefore assume that
inflation happened at an intermediate
scale, below the GUT scale.

We shall now discuss such an example of an intermediate scale supersymmetric hybrid inflation model which solves
the $\mu$ problem and the strong CP problem \cite{Bastero-Gil:1997vn,Bastero-Gil:1999fz,Bastero-Gil:2000jd}.
The model is based on the superpotential:
\beq
W= \lambda N H_u H_d - \kappa \phi N^2 \,,
\label{superpot}
\eeq
where $N$ and $\phi$ are singlet superfields, and $H_{u,d}$ are the Higgs
superfields, and $\lambda,\kappa$ are dimensionless couplings.
Other cubic terms in the superpotential are forbidding by imposing a
global $U(1)_{PQ}$ Peccei-Quinn symmetry.
The superpotential in Eq. (\ref{superpot}) includes a linear
superpotential for the inflaton field, $\phi$, typical of hybrid inflation,
as well as the singlet $N$ coupling to Higgs doublets as in the NMSSM.
In the original version of this model we assumed that
during inflation $N$ and $H_u$, $H_d$ were set to zero, so that the inflationary trajectory was
as in Figure \ref{infold} due to the potential
\bea
V&=&V(0)+ \frac{\kappa^2}{4}N^4 + \kappa^2(\phi-\phi_c^+)(\phi-\phi_c^-)N^2
+ \frac{1}{2} m_\phi^2 \phi^2
\label{V}
\eea
The critical value of $\phi$  corresponds to the field dependent mass squared
$\kappa^2(\phi-\phi_c^+)(\phi-\phi_c^-)N^2$ changing sign and becoming
negative. At the global minimum (NOW in Figure \ref{infold}) $\mu=\lambda<N> \sim 1 TeV$,
and the model incorporates an axionic solution to the strong CP problem with
$f_a\sim <N>\sim <\phi>\sim 10^{13}GeV$. Of course this leads to a naturalness problem
since it requires $\lambda \sim \kappa \sim 10^{-10}$. The model also leads to
$V_0^{1/4}\sim 10^8 GeV$ and $H_*\sim 10MeV$. Slow roll only requires $m_{\phi}\sim MeV$
but the COBE constraint requires $m_{\phi}\sim eV$! This is an example of how the
requirement of structure formation arising from the inflaton imposes severe constraints on the theory,
and motivates an application of the curvaton approach to this model.

Recently \cite{Bastero-Gil:2002xr}
an alternative inflationary trajectory was discussed
in which these fields may take small values away from the origin,
consistent with slow roll inflation. The motivation for considering such a trajectory
was to allow the Higgs fields to slowly roll during inflation, and so play the
r\^{o}le of the curvaton. However, the Higgs fields are not late decaying scalars since
they have gauge couplings which ensure rapid decay, so the mechanism which allows
the Higgs fields to be responsible for large scale structure cannot be the curvaton
mechanism as it was originally envisaged
\cite{Lyth:2001nq,Lyth:2002my,Moroi:2001ct,Moroi:2002rd,Dimopoulos:2002kt,Dimopoulos:2003ii,Dimopoulos:2003az}.
The new mechanism of structure formation that we proposed \cite{Bastero-Gil:2002xr}
relies on the observation that in hybrid inflation at the start of reheating,
the vacuum energy present during inflation $V(0)$ gets
redistributed among all the oscillating fields such that their energy densities become
comparable. Thus any isocurvature perturbation in one of the
hybrid inflation fields (in this case the Higgs field) may be converted into curvature perturbations
during the on-set of reheating. In such a scenario, the usual Higgs field responsible for
the origin of mass in the supersymmetric standard model could also be responsible
for generating the large scale structure in the universe!

In order to satisfy the D-flatness of the new inflationary trajectory
we assumed \cite{Bastero-Gil:2002xr}
the values of the Higgs doublets during inflation to
be equal, $H_u=H_d=h$. An important condition for inflation is that
the inflaton mass $m_\phi$ (and also $m_h$) needs to
be small enough in order to ensure the slow-roll of the inflaton.
This implies $h\ll \phi$ and $\dot{h}\ll \dot{\phi}$.
During inflation the curvature dominated by $\phi$ is very much less than
the entropy which is dominated by $h$,
\beq
{\cal R}\sim -\frac{H_*}{\dot{\phi}_*}\delta \phi_*\ll
S\sim -\frac{H_*}{\dot{h}_*}\delta h_*
\label{RllS}
\eeq
The isocurvature Higgs perturbations are then transferred to the curvature
perturbations at the start of reheating when the energy densities
become comparable, $\rho_{\phi}\sim \rho_h$, and the numerical solution
to the field equations leads to a resulting curvature ${\cal R}\sim 0.1S$ \cite{Bastero-Gil:2002xr}.
The model gives non-Gaussianity below the Planck sensitivity,
and a spectral index differing from unity by a number of order $0.1$.

The above mechanism for structure formation allows $m_{\phi}\sim MeV$ rather than $eV$,
which alleviates but does not solve all the naturalness problems of the model.
However a completely natural model is possible using extra dimensions to give
volume suppression factors for masses and couplings \cite{Bastero-Gil:2002xs}.
We suppose the Higgs and singlets $N,\phi$ are in an extra-dimensional
bulk, and the matter fields live on our brane and
feel the usual 3+1 dimensions. We set the string scale $M_*\sim
10^{13} GeV$ and the supersymmetry breaking scale
$\sqrt F_S\sim 10^8 GeV$. Then we find natural values for all physical
quantities. The vacuum energy is as required
$V(0)^{1/4}\sim \sqrt F_S\sim 10^8 GeV$.
The small couplings are also as required
$\lambda , \kappa \sim (M_*/m_P)^2\sim 10^{-10}$.
The soft masses on the supersymmetry breaking brane are
$m_{soft}\sim F_S/M_*\sim TeV$,
but the soft masses for bulk scalars are
$m_{\phi}\sim m_{h}\sim m_{N}\sim (M_*/m_P)m_{soft}\sim MeV$,
which satisfy the slow roll conditions.

The complete model \cite{Bastero-Gil:2002xr,Bastero-Gil:2002xs}
is therefore completely natural, and accounts for
large scale structure via quantum fluctuations of the Higgs field
during inflation, which are subsequently
transferred to curvature perturbations at the on-set of reheating.

\section{Conclusion}

These are exciting times for inflationary cosmology due to the new data
from WMAP, and the prospect of the Planck launch in 2007,
which will extend the angular power spectrum out to $l\sim 3000$, measure the polarization,
measure the flatness to an accuracy
$\Delta \Omega_0\approx \pm 0.001$ and the spectral index
to $\Delta n_s\approx \pm 0.008$.
Inflation is already looking very good,
with hybrid inflation at the intermediate scale looking most relevant for particle physics,
and providing a link with the supersymmetric standard model \cite{Kane:2001rb}.
Curvature perturbations need not have arisen from the inflaton,
and could have originated from a curvaton or Higgs.

\begin{center}
{\bf Acknowledgment}
\end{center}
I would like to thank my collaborators Mar Bastero-Gil and Vicente Di
Clemente, and also the local organisers of PASCOS'03 for making it
such a highly successful and enjoyable conference.

\end{document}